\documentclass{mem}
\usepackage{natbib}\usepackage{txfonts}\usepackage{balance}
\usepackage{graphicx}
\usepackage{txfonts}
\usepackage[a4paper]{hyperref}
\idline{79}{3}
\begin{document}
\def\teff{$T\rm_{eff }$}
\def\kms{$\mathrm {km s}^{-1}$}

\title{Pulsational models of BL Her stars 
}

   \subtitle{}

\author{ 
M. \,Di Criscienzo$^3$,
M. \, Marconi$^1$,
F.\, Caputo$^3$,
S.\, Cassisi$^2$,
          }
\offprints{M. Di Criscienzo}
\institute{
$^1$INAF-Osservatorio Astronomico di Capodimonte, Napoli, Italy
 \email: dicrisci@na.astro.it\\
$^2$INAF-Osservatorio Astronomico di Teramo, Italy\\
$^3$INAF- Osservatorio Astronomico di Roma, Monteporzio Catone,(RM) Italy}

\authorrunning{Di Criscienzo}

\titlerunning{Pulsational models of BL Her stars}

\abstract{
In this paper we present an updated and homogeneous
pulsational scenario
for a wide range of stellar parameters typical of BL Her stars i.e.,
Population II Cepheids with periods shorter than 8 days.

}
\maketitle{}

\section{Introduction}
Population II pulsating variables play a fundamental role in our understanding of the properties of old stellar populations as well as in the definition of the cosmic distance scale. Among them, Population II Cepheids (P2Cs), with periods P from 1 to 5 days, are observed in clusters with few RR Lyrae stars and blue Horizontal Branch (HB) morphology. These variables are brighter than RR Lyrae stars with similar metal content and  are often separated into BL Her stars (logP $<$ 1) and W Vir stars (logP $>$ 1). As reviewed in  Wallerstein (2002) they originate from hot, low-mass stellar structures that started the main central He-burning phase in the blue side of the RR Lyrae gap and now evolve toward the AGB crossing the pulsation region with luminosity and effective temperature that increase with decreasing the mass.\\
Following our  program devoted to a homogeneous study of radially pulsating stars with various chemical  compositions, masses and luminosities, here we discuss the results of updated pulsation models with mass 0.50-0.65 $M_{\odot}$ and luminosity LogL/$L_{\odot}$=1.81 - 2.41, in order to build a sound theoretical scenario for the analysis of the short period P2Cs.
\section{Models \& results}
\begin{table}
\caption{Input parameters of the computed BL Her models. A 
helium abundance Y=0.24 has been adopted.}                
\label{table:1}      
\centering   
\setlength{\tabcolsep}{0.01in}
\begin{tabular}{c c c c c c c}        
\hline\hline                 
Z& M/M$_{\odot}$ & LogL/L$_{\odot}$& FOBE&FBE&FORE&FRE \\    
\hline                        
0.0001& 0.60 & 1.95 &-&6850&-&5750\\
      &      & 2.05 &-&6750&-&5650\\
      &      & 2.15 &-&6750&-&5550\\     
      & 0.65 & 1.91 &6950&6850&6100&5750\\
      &      & 2.01 &6750&6850&6300&5750\\
      &      & 2.11 &-&6750&-&5550\\
\hline
0.001 & 0.50 & 2.11 &-&6650&-&5450\\
      &      & 2.41 &-&6350&-&5150\\
      & 0.55 & 1.81 &6875&6850&6400&5650\\
      &      & 1.91 &-&6850&-&5550\\
      &      & 2.01 &-&6750&-&5450\\
      & 0.65 & 1.81 &7050&6750&6650&5750\\
      &      & 1.91 &6850&6750&6150&5650\\
      &      & 2.01 &6650&6850&6350&5650\\
\hline
0.004 & 0.55 & 1.81 &-&6950&-&5750 \\
      &      & 1.91 &-&6850&-&5650 \\
      &      & 2.01 &-&6750&-&5450 \\
\hline                                   
\end{tabular}
\end{table}
The pulsation models, listed in Table 1, were calculated using the nonlinear, nonlocal and time-dependent convective hydrodynamical code of Bono \& Stellingwerf (1994) with the same physical assumptions (i.e., equation of state and opacity tables) already used for the analysis of  RR Lyrae stars (see Di Criscienzo et al. 2004 and references therein).\\
The model sequences were computed as one parameter families with  constant chemical composition, mass and luminosity, by varying the effective temperature Te by steps of 100 K. In these computations we adopted a value of the mixing length parameter l/Hp=1.5 to close the system of convective and dynamical equations (see Marconi \& Di Criscienzo 2007 for details). Here we present some relevant results.  \\
\begin{figure}[b]
\resizebox{\hsize}{!}{\includegraphics[clip=true]{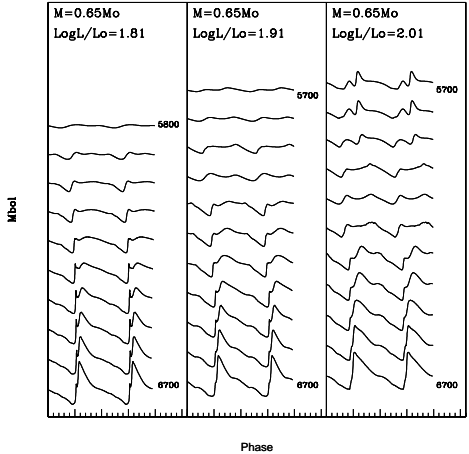}}
\caption{Theoretical bolometric light curves
               for a subsample (Z=0.001 and M=0.65 M$_\odot$) of fundamental models.The luminosity levels are labeled.}
\end{figure}
Starting with the models with 0.65 M$_\odot$ and logL/L$_\odot$=1.81, we note that they follow the well known behaviour of RR Lyrae stars in the instability strip, with FO models  generally bluer than the F ones, but  the FO-Red Edge (RE) redder than the F-Blue Edge (BE) (see Table 1). As a consequence, we have that:  
\begin{itemize}
\item the limits of the whole pulsation region are described by the FOBE and the FRE;
\item both the pulsation modes are stable in the middle zone delimited by the F-BE and tvhe FORE;
\item F-only pulsators are located between the FRE and the FORE and FO pulsators between the FBE and the FOBE.
\end{itemize}
By increasing the luminosity, the whole pulsation region moves towards the red, but with a significant shrinking of the FO-only pulsation region. As one can see in Table 1 the difference in effective temperature between the FOBE and the FBE is about 300 K at LogL/$L_{\odot}$=1.81 and about 100 K at LogL/$L_{\odot}$=1.91. A further increase of the luminosity yields that the FOBE becomes redder than the FBE with the total disappearance of stable FO models at LogL/L$L_{\odot}$$\ge$2.11.\\
Varying the mass, we note that with 0.60$M_{\odot}$ no FO model is stable at Log/$L_{\odot}$$\ge$ 1.95, while with 0.55$M_{\odot}$ we get that the FOBE coincides with the FBE at LogL/$L_{\odot}$=1.80 and only F models are present above this luminosity level.\\
 In summary, the results listed in Table 1 confirm earlier suggestions  that for each given mass and helium content  there exists an  ``intersection'' luminosity LIP where the FOBE intersects the FBE, and that above this luminosity only the fundamental mode is stable. On this ground, one has that the red limit of the instability strip is always determined by the FRE, while the blue limit is given by the FO-BE or the F-BE depending on whether the luminosity is fainter or brighter, respectively, than LIP. \\
Moreover, we wish to recall that the onset of pulsation depends also on the efficiency of convection in the star external layers, namely on the adopted value of the mixing length parameter l/Hp. Since the effect of convection is to quench pulsation and the depth of convection increases from high to low effective temperatures, we expect that varying the l/Hp value will modify the effective temperature at the FRE by a larger amount with respect to the FBE or the FOBE. Indeed, the additional computations with l/Hp=2.0 computed by Marconi \& Di Criscienzo (2007) have confirmed the general trend shown by RR Lyrae (Marconi et al. 2003), with the FBE and FRE effective temperatures increasing by about 100 and about 300 K, respectively, at constant mass and luminosity.\\
For each investigate model, the non linear analysis provides the variation of relevant parameters, namely luminosity, radius, radial velocity, effective temprerature and gravity along a pulsational cycle. A subsample of bolometric curves is shown in Fig.1. All these curves show a large variety of shapes, which is perhaps the most strinking feature of BL Her models. \\
To compare theoretical results with observations, the bolometric light curves have been transformed into the photometric bands UBVRIJK usig bolometric corrections and temperature color relations provided by Castelli et al. (1997,a,b). In this way, light- curve amplitudes and mean absolute magnitudes are derived in every photometric bands. As an example, in Fig.2 we report the behaviour of visual amplitudes of fundamental models as a function of the period for different values of metallicity, mass and luminosity. Similarly to what happens for RR Lyrae, we find that for a fixed period the amplitude increases as the luminosity increases and the mass decreases, while it remains quite constant in the considered range of metallicity.\\
It is important to note that part of the observed behaviour in the period-amplitude diagram is due to the dependence of  period on mass and luminosity. The change in the pulsation amplitude in mainly related to the distance of the FBE, as shown in the two right panels of the same figure. The deviation from the  linearity of the highest luminosity models is related to the complex coupling between pulsation  and convection for these low density and cool structures. However, in the range of linearity we obtain a relation  in agreement with that calculated for RR lyrae stars in Di Criscienzo et al. (2004). 
\begin{figure}[]
\resizebox{\hsize}{!}{\includegraphics[clip=true]{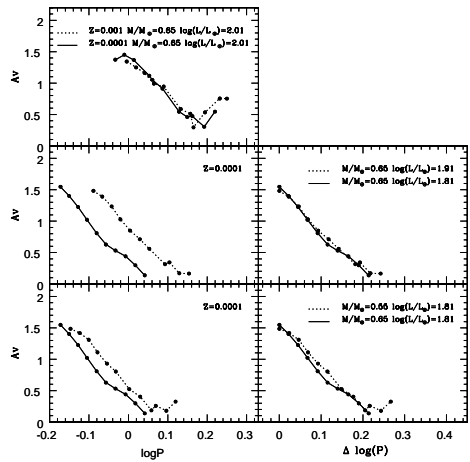}}
\caption{Left panels: visual amplitudes versus periods for selected fundamental models  varying the metallicity at fixed mass and luminosity (top), 
   varying the luminosity at fixed mass and 
metallicity (Z=0.001, middle) and varying the mass at fixed luminosity and 
metallicity (Z=0.001, bottom). Right panels: visual amplitudes plotted versus 
${\Delta}$ log(P)=log (P)- log(P$_{FBE}$) varying the luminosity  at fixed mass and 
metallicity and varying the mass at fixed luminosity and 
metallicity.
    }
\label{li_vhel}
\end{figure}

\bibliographystyle{aa}

\end{document}